\documentclass[conference,10pt]{IEEEtran}
\IEEEoverridecommandlockouts
\usepackage{citesort}
\usepackage{amsmath,amssymb,amsfonts}
\usepackage[shortlabels]{enumitem}
\usepackage[ruled]{algorithm2e}
\usepackage{algorithmic}
\usepackage{graphicx}
\usepackage{textcomp}
\usepackage{xcolor}
\usepackage{epsfig}
\usepackage{subfigure}
\usepackage{psfrag}
\def\BibTeX{{\rm B\kern-.05em{\sc i\kern-.025em b}\kern-.08em
		T\kern-.1667em\lower.7ex\hbox{E}\kern-.125emX}}

\begin{document}
\title{Decentralized Federated Learning via SGD over Wireless D2D Networks}
\author{\IEEEauthorblockN{Hong Xing${}^{\ast\S}$, Osvaldo Simeone${}^\S$, and Suzhi Bi${}^\ast$}\\
	\IEEEauthorblockA{${}^\ast$College of Information Engineering, Shenzhen University, Shenzhen, China\\
		${}^\S$KCLIP Lab, CTR, Department of Engineering, King's College London, London, U.K.\\
		E-mails:~hong.xing@szu.edu.cn,~osvaldo.simeone@kcl.ac.uk,~bsz@szu.edu.cn}
}
\maketitle
\begin{abstract}
Federated Learning (FL), an emerging paradigm for fast intelligent acquisition at the network edge, enables joint training of a machine learning model over distributed data sets and computing resources with limited disclosure of local data. Communication is a critical enabler of large-scale FL due to significant amount of model information exchanged among edge devices. In this paper, we consider a network of wireless devices sharing a common fading wireless channel for the deployment of FL. Each device holds a generally distinct training set, and communication typically takes place in a Device-to-Device (D2D) manner. In the ideal case in which all devices within communication range can communicate simultaneously and noiselessly, a standard protocol that is guaranteed to converge to an optimal solution of the global empirical risk minimization problem under convexity and connectivity assumptions is Decentralized Stochastic Gradient Descent (DSGD). DSGD integrates local SGD steps with periodic consensus averages that require communication between neighboring devices. In this paper, wireless protocols are proposed that implement DSGD by accounting for the presence of path loss, fading, blockages, and mutual interference. The proposed protocols are based on graph coloring for scheduling and on both digital and analog transmission strategies at the physical layer, with the latter leveraging over-the-air computing via sparsity-based recovery. 
\end{abstract}


\IEEEpeerreviewmaketitle
\newtheorem{definition}{\underline{Definition}}[section]
\newtheorem{fact}{Fact}
\newtheorem{assumption}{Assumption}
\newtheorem{theorem}{\underline{Theorem}}[section]
\newtheorem{lemma}{\underline{Lemma}}[section]
\newtheorem{proposition}{\underline{Proposition}}[section]
\newtheorem{corollary}[proposition]{\underline{Corollary}}
\newtheorem{example}{\underline{Example}}[section]
\newtheorem{remark}{\underline{Remark}}[section]
\newcommand{\mv}[1]{\mbox{\boldmath{$ #1 $}}}
\newcommand{\mb}[1]{\mathbb{#1}}
\newcommand{\Myfrac}[2]{\ensuremath{#1\mathord{\left/\right.\kern-\nulldelimiterspace}#2}}
\newcommand\Perms[2]{\tensor[^{#2}]P{_{#1}}}

\section{Introduction}
Distributed learning refers to scenarios with multiple agents collaboratively training a machine learning model over geographically distributed computing resources and data, with examples ranging from decentralized data centers to Internet-of-Things (IoT) networks \cite{savazzi20decentralized,chang20distributed}.  
Communication is a critical enabler of distributed learning systems \cite{bekkerman11scaling}, and effective collaborative training often involves the exchange of significant amounts of information, including training data, model parameters, and gradient vectors.
\begin{figure}[tp]
	\centering
	\subfigure[A connectivity graph for a D2D wireless network with path loss, fading and blockages. \label{subfig:illustration of vertex coloring}]{\includegraphics[width=2.4in]{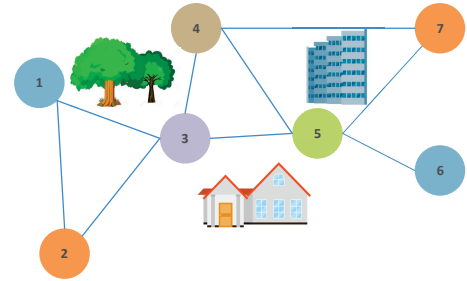}}
	\subfigure[Timeline of iterations with multi-slotted communication blocks used to exchange model information.\label{subfig:illustration of communication blocks}]{\includegraphics[width=2.8in]{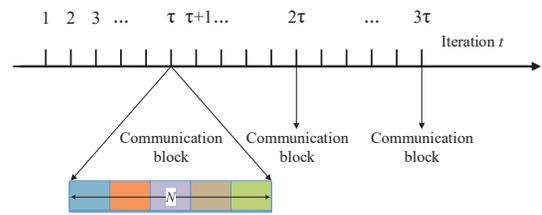}}
	\caption{Decentralized federated learning over wireless D2D networks.}\label{fig:system model}
	\vspace{-.2in}
\end{figure}

Agents in distributed learning systems are typically connected either in a star topology, such as in a master-worker architecture, or in a decentralized device-to-device (D2D) topology characterized by a general graph (see example in Fig. \ref{fig:system model}). 
For star topologies, {\it federated learning (FL)} \cite{kairouz19advances} has recently been widely studied as a means to enable cooperative training based on the exchange of model information through the center node. 
For D2D topologies, devices can only communicate with their neighbors, and consensus mechanisms are needed to ensure the achievement of a common learning goal. {\color{black} A well-known solution to large-scale machine learning problems in D2D topologies with ideal communication is {\it Decentralized Stochastic Gradient Descent (DSGD)}, which guarantees convergence to optimality under assumptions of convexity and connectivity \cite{ram10distributed}. Variants of DSGD have been developed to improve its convergence performance, including gradient tracking algorithms \cite{xin19decentralized}. For graphs with interference constraints,  matching-based multi-user scheduling schemes have been proposed in \cite{wang19MATCHA}, enabling parallel inter-node information exchange .} Communication overhead for distributed learning can also be reduced via sparsification and/or quantization \cite{basu19qsparse}, possibly combined with event-triggered mechanisms  \cite{singh19sparq-sgd}.

{\color{black} All the prior works summarized above assume noiseless or rate-limited communications, hence not accounting for wireless impairments caused by link blockages, channel fading, and mutual interference.} For star topologies, FL in wireless networks was studied in several recent works. In \cite{chen19joint}, a joint learning, wireless resource allocation, and user selection problem was formulated to mitigate the impact of wireless impairments. \emph{over-the-air computation (AirComp)} was investigated in \cite{Gunduz19over-the-air,zhu20one-bit,ahn19distillation} as a promising solution to support simultaneous  transmissions in large-scale FL by leveraging the superposition property of the wireless medium. 

{\color{black}In this work, we study for the first time the problem of implementing DSGD for server-less edge networks. The contribution of this paper is summarized as follows. 1) We propose protocols for the implementation of DSGD over wireless D2D networks by considering both digital and analog transmission schemes, with the analog implementation leveraging over-the-air computing. 2) To cope with wireless interference, we apply graph-coloring based scheduling strategies to the designs of digital and analog implementations. 3) Simulations are presented to benchmark the proposed digital and analog implementations against various baseline strategies, providing insights on the performance comparison between digital and analog implementations.}  

The remainder of this paper is organized as follows. The learning problem and the communication models are presented in Section \ref{sec:System Model}. The digital and analog implementations of the DSGD algorithm are, respectively, introduced in Sections \ref{sec:Digital Implementation} and \ref{sec:Analogue Implementation}. The simulation results are provided in Section \ref{sec:Experiments}. Finally, Section \ref{sec:Conclusion and Discussions} concludes the paper.


\section{System Model}\label{sec:System Model} 
In this paper, as illustrated in Fig. \ref{fig:system model}, we consider a decentralized {federated} learning model, in which a set \(\mathcal{V}=\{1,\ldots,K\}\) of $K$ devices, each with a local training data set, collaboratively train a shared machine learning model via wireless D2D communications without directly disclosing their respective training samples to each other \cite{kairouz19advances}.

\subsection{Data and Learning Model}
Each device \(i\in\mathcal{V}\) has a data set \(\mathcal{D}_i\), where different sets \(\mathcal{D}_i\) and \(\mathcal{D}_j\), \(i\neq j\), possibly have non-empty intersections. All devices share a common machine learning model, e.g., a class of neural networks, which is parametrized by a vector \(\mv\theta\in\mb{R}^{d\times 1}\). The goal of the network is to collaboratively tackle the empirical loss minimization problem
\begin{align*}
\mathrm{(P0)}:&~\mathop{\mathtt{Minimize}}_{\mv \theta}~~~F(\mv\theta)\triangleq\frac{1}{K}\sum_{i\in\mathcal{V}}f_i(\mv \theta),
\end{align*} 
where \(F(\mv \theta)\) is the global empirical loss function;  \(f_i(\mv\theta)=\tfrac{1}{\vert\mathcal{D}_i\vert}\sum_{\mv\xi\in\mathcal{D}_i}l(\mv\theta,\mv\xi)\) is the local empirical loss function for the data available at device $i$; and \(l(\mv\theta,\mv\xi)\) denotes the loss function (e.g., cross entropy for classification problems) for the parameter vector \(\mv\theta\) evaluated on a data sample \(\mv\xi\).
At any $t$th iteration of the distributed learning process, each device $i\in\mathcal{V}$ has a local parameter vector \(\mv\theta_i^{(t)}\) that approximates the solution to problem (P0).


In this paper, we adopt the standard decentralized stochastic gradient descent (DSGD) algorithm \cite{ram10distributed,xin19decentralized}, suitably modified to account for constraints arising from wireless transmission. In conventional DSGD, the devices are assumed to be connected over an undirected graph \(\mathcal{G}(\mathcal{V},\mathcal{E})\), with \(\mathcal{V}\) denoting the set of nodes and \(\mathcal{E}\subseteq\{(i,j)\in\mathcal{V}\times\mathcal{V}\left.\vert\right.i\neq j\}\) the set of edges. The devices carry out local SGD updates and communicate periodically every $\tau$ iterations. As a result, at each iteration $t$ that is not a multiple of $\tau$, device $i\in\mathcal{V}$ executes an SGD step based on its data set \(\mathcal{D}_i\) by updating its local parameter \(\mv\theta_i^{(t)}\) as
\begin{align}
\mv\theta_i^{(t)}=\mv\theta_i^{(t-1)}-\eta^{(t)}\hat\nabla f_i(\mv\theta_i^{(t-1)}), \label{eq:local updates}
\end{align}  
where \(\eta^{(t)}\) denotes the learning rate, possibly dependent on $t$; and \(\hat\nabla f_i(\mv\theta_i^{(t-1)})\) is the estimate of the gradient \(\nabla f_i(\mv\theta)\) obtained from a mini-batch \(\mathcal{D}_i^{(t)}\subseteq\mathcal{D}_i\) of device $i$'s data samples as 
\begin{align}
\hat\nabla f_i(\mv\theta_i^{(t-1)})=\frac{1}{\vert\mathcal{D}_i^{(t)}\vert}\sum\limits_{\mv\xi\in\mathcal{D}_i^{(t)}}\nabla l(\mv\theta_i^{(t-1)},\mv\xi). \label{eq:SGD}
\end{align}
In contrast, when $t$ is an integer multiple of $\tau$, the nodes exchange their current local parameters with their neighbors in graph $\mathcal{G}$ and perform a \emph{consensus update}. Mathematically, each device $i\in\mathcal{V}$ linearly combines the parameters received from the set \(\mathcal{N}_i\) of neighbors in graph $\mathcal{G}$ with weights \(\{w_{ij}\}_{j\in\mathcal{N}_i}\) along with the local SGD update as 
\begin{align}
\mv\theta_i^{(t)}=w_{ii}\mv\theta_i^{(t-1)}+\sum\limits_{j\in\mathcal{N}_i}w_{ij}\mv{\theta}_{j}^{(t-1)}-\eta^{(t)}\hat\nabla f_i(\mv\theta_i^{(t-1)}). \label{eq:consensus udpates}
\end{align}

{\color{black}We consider the standard choice of weight matrix \([\mv W]_{ij}\triangleq w_{ij}\) with \(w_{ij}=\alpha\), \(\forall j\in\mathcal{N}_i\), \(w_{ii}=1-\vert\mathcal{N}_i\vert\alpha\), and \(0\) otherwise, \(i\in\mathcal{V}\),
where the constant \(\alpha\) is a design parameter related to the topology of graph $\mathcal{G}$ to ensure fast consensus \cite{xiao04fast}.} A typical choice is \(\alpha=\Myfrac{2}{(\lambda_1(\mv L)+\lambda_{K-1}(\mv L))}\), where \(\mv L = \mv D - \mv A\) is the Laplacian of the graph, with \(\mv D\) and \(\mv A\) denoting the degree matrix and the adjacency matrix of the graph, respectively, and \(\lambda_i(\cdot)\) denotes the $i$th largest eigenvalue of its argument matrix. 


\subsection{Communication Model}

For the entire duration of the collaborative learning session of interest, communication between any two nodes $i$ and $j$ may be blocked due to shadowing with probability $p_{ij}$ independently of all other pairs. The unblocked links thus define  
a connectivity graph \(\mathcal{G}(\mathcal{V},\mathcal{E})\), with \(\mathcal{E}\subseteq\{(i,j)\in\mathcal{V}\times\mathcal{V}\left.\vert\right.i\neq j\}\) denoting the edge set for the unblocked pairs. 
Devices $i$ and $j$ are referred to as {\it connected} if edge ($i,j$) is included in the connectivity graph $\mathcal{G}$. 

As seen in Fig. \ref{subfig:illustration of communication blocks}, iteration \(t=\tau, 2\tau, \ldots\) of the learning algorithm is associated with a communication block of $N$ channel uses. The signal received at device $i\in\mathcal{V}$ during a channel use $n=1,\ldots, N$ for the $t$th iteration (\(t=\tau, 2\tau, \ldots\)) can be written as  
\begin{align}
y_{i,n}^{(t)}=\sum\limits_{j\in\mathcal{N}_i}\sqrt{A_0}\left(\frac{d_0}{d_{ij}}\right)^{\Myfrac{\gamma}{2}}h_{ji}^{(t)}x_{i,n}^{(t)}+n_{i,n}^{(t)}, \label{eq:received signal}
\end{align} where \(h_{ij}^{(t)}\sim\mathcal{CN}(0,1)\) is the Rayleigh fading channel coefficient between device $i$ and device $j$ at the $t$th iteration, which is assumed to vary independently across communication blocks; \({n}_{i,n}^{(t)}\sim\mathcal{CN}(0,N_0)\) is the additive white Gaussian noise (AWGN) at device $i$; \(x_{i,n}^{(t)}\) is the signal transmitted by device $i$; \(A_0\) is the average channel gain at reference distance $d_0$; \(d_{ij}\) is the distance between device $i$ and $j$; and \(\gamma\) is the path loss exponent factor. An average power constraint of \(\Myfrac{1}{N}\sum_{n=1}^N\mb E[\vert x_{i,n}^{(t)}\vert^2]\le\bar P\) is imposed for all devices \(i\in\mathcal{V}\)  and communication blocks \cite{amiri19fading}. We assume channel state information about the local channels \(\{h_{ij}\}_{j\in\mathcal{N}_i}\) to be available at each device \(i\in\mathcal{V}\). 


\section{Digital Implementation}\label{sec:Digital Implementation}
In this section, we propose a digital implementation of the DSGD algorithm by detailing the scheduling and physical-layer transmission policies.

\subsection{Scheduling}\label{subsec:digital scheduling}
Each communication block at iteration \(t=\tau,2\tau,\ldots\) (see Fig. \ref{subfig:illustration of communication blocks}) is divided into $M$ time slots, and a different subset of devices transmit a quantized version of the respective local parameter vectors \(\mv\theta_i^{(t)}\) in each slot. The subsets of devices are scheduled so that: {\it(i)} no two connected devices are scheduled to transmit in the same slot due to the half-duplex transmission constraint; and {\it(ii)} no two devices connected to the same device are scheduled to transmit in the same slot, so as not to cause interference at a common neighbor. 

To elaborate, we construct an auxiliary graph \(\mathcal{G}^d(\mathcal{V},\mathcal{E}^d)\) such that the edge set \(\mathcal{E}^d\supseteq\mathcal{E}\) includes not only the original edges in \(\mathcal{E}\), but also one edge for each pair of nodes sharing one or more common neighbors. We then carry out vertex coloring on the auxiliary graph \(\mathcal{G}^d(\mathcal{V},\mathcal{E}^d)\), such that any two nodes connected by an edge are assigned with distinct ``colors''. The minimum number of colors required is the {\em chromatic number} of the graph. Finally, scheduling proceeds by assigning the nodes with the same ``color'' to the same slot (see Fig. \ref{subfig:illustration of vertex coloring}). This ensures that both requirements {\it(i)} and {\it(ii)} above are satisfied. Since finding an optimal vertex coloring for a general graph is known to be NP-hard, we adopt the well-known greedy heuristic algorithm. The greedy algorithm has linear-time complexity, and yields a number of colors that is upper-bounded by the maximum degree of the graph \cite[Sec. 2]{Husfeldt15graph}. For the proposed digital implementation, the number $M$ of time slots is henceforth given by the number of colors obtained by the greedy algorithm. 

\subsection{Physical-Layer Transmission}
As a result of the the scheduling strategy designed above, each device $i\in\mathcal{V}$ transmits in a single time slot, and there is no interference among simultaneous transmissions. Each slot contains \(\lfloor\Myfrac{N}{M}\rfloor\) channel uses. In the scheduled transmission slot, the device broadcasts to its neighbors a quantized version of its local parameter \(\mv\theta_i^{(t)}\). The number \(B_i^{(t)}\) of bits that device $i\in\mathcal{V}$ can successfully transmit to its neighbors during a time slot \(t=\tau, 2\tau, \ldots\) is limited by the neighbor with the lowest data rate, i.e., 
\begin{align}
B_i^{(t)} = \left\lfloor\frac{N}{M}\right\rfloor\log_2\left(1+\frac{\bar PM}{N_0}\min\limits_{j\in\mathcal{N}_i}\left\{A_0\left(\frac{d_0}{d_{ij}}\right)^\gamma\vert h_{ij}^{(t)} \vert^2\right\}\right). \label{eq:transmission bits for D-DSGD}
\end{align} 
In order to quantize the local parameter vector \(\mv\theta_i^{(t)}\) to \(B_i^{(t)}\) bits, each device $i\in\mathcal{V}$ applies a compression operation  \(\mathrm{comp}_{B_i^{(t)}}(\cdot)\) that composes a sparsifying operation with deterministic quantization. This compression operator is applied to an error-compensated parameter \(\mv\theta_i^{(t)} + \mv e_i^{(t-1)}\), where \(\mv e_i^{(t)}\) with \(\mv e_i^{(0)} = \mv 0\) denotes the accumulated error for device $i$ at the $t$th iteration, which is updated as 
\begin{align}
\mv e_i^{(t)}=\mv e_i^{(t-1)}+\left(\mv\theta_i^{(t)}
-\mathrm{comp}_{B_i^{(t)}}(\mv\theta_i^{(t)}+\mv e_i^{(t-1)})\right).\label{eq:accumulated quantization error}
\end{align} 

The sparsifying operation sets all but the $l$ elements of \(\mv\theta_i^{(t)}+\mv e_i^{(t-1)}\) with the largest absolute to zero. 
This is followed by a quantization scheme that employs a number of \(\log_2\binom{d}{l}\) bits for encoding the position of the non-zero elements in vector \(\mv\theta_i^{(t)}+\mv e_i^{(t-1)}\) after sparsification, while $b$ bits are used to quantize the value of each of the $l$ elements. Parameter $l$ is chosen as the maximum value that satisfies the bit budget constraint
\begin{align}
\log_2\binom{d}{l} + bl \le B_i^{(t)}. \label{eq:maximum permissive sparse level}
\end{align} 

After each device $i\in\mathcal{V}$ receives the parameters one by one from its neighboring set \(\mathcal{N}_i\), at the end of iteration \(t=\tau,2\tau,\ldots\), it performs the consensus update as (cf.~\eqref{eq:consensus udpates})
\begin{multline}
\mv\theta_i^{(t+1)}=w_{ii}\mv\theta_i^{(t)}+\sum\limits_{j\in\mathcal{N}_i}w_{ij}\mathrm{comp}_{B_j^{(t)}}\left(\mv\theta_j^{(t)}+\mv e_j^{(t-1)}\right)-\\\eta^{(t)}\hat\nabla f_i(\mv\theta_i^{(t-1)}). \label{eq:consensus udpates for D-DSGD}
\end{multline}

\section{Analog Implementation} \label{sec:Analogue Implementation}
In this section, we propose an analog implementation of the DSGD algorithm based on over-the-air computing by detailing scheduling and physical-layer strategies. 

\subsection{Scheduling}
Existing over-the-air computing techniques operate over star topologies {\color{black}in a pair of consecutive time slots} \cite{Gunduz19over-the-air,zhu20one-bit}. In the first slot, the node $i$ at the center (e.g., parameter server in an FL system) receives a superposition of the signals simultaneously transmitted by all its neighbors in \(\mathcal{N}_i\) (serving as AirComp transmitters). From this noisy observation, the center node $i$ estimates the sum of parameter vectors \(\sum_{j\in\mathcal{N}_i}w_{ij}\mv\theta_j\) in \eqref{eq:consensus udpates}. In the second slot, the center node $i$ broadcasts its updated parameter \eqref{eq:consensus udpates} to all its neighbors in \(\mathcal{N}_i\) (serving as broadcasting transmitters). In order to leverage over-the-air computing in a D2D topology, we propose a scheduling policy that aims at selecting as many non-interfering subnetworks with star topologies as possible in each pair of time slots. 
%

{\color{black} To this end, we first carry out the greedy coloring algorithm described in Section \ref{subsec:digital scheduling} on graph \(\mathcal{G}^{(1)}=\mathcal{G}(\mathcal{V},\mathcal{E})\). We define \(d_c^{(1)}\) as the sum of the degrees of all nodes that have been assigned the same color $c$. Next, we schedule all nodes assigned the degree-maximizing color $c^\ast = \arg\max\{d_c^{(1)}\}$  as ``star centres''. In the first slot, the scheduled nodes receive combined signals from all their neighbors in \(\mathcal{G}^{(1)}\). In the second slot, the scheduled nodes broadcast to all their neighbors in \(\mathcal{G}^{(1)}\). The scheduled nodes and their connected edges, along with any nodes disconnected from \(\mathcal{G}^{(1)}\), are then removed from the graph to produce a residual graph \(\mathcal{G}^{(2)}\). The procedure outlined above, including the coloring step, is repeated on the residual graph \(\mathcal{G}^{(2)}\) to schedule transmissions in slots $3$ and $4$. The overall procedure is repeated for all successive pairs of slots, until the residual graph is empty.}

\subsection{Physical-Layer Transmissions}
Unlike the digital implementation, each device generally transmits in multiple slots (any time a neighbour is scheduled as a star center or itself is scheduled as a broadcasting transmitter), but it only acts as an over-the-air computing receiver once in one communication block. We denote as \(n_i^c\le\vert\mathcal{N}_i\vert\) the number of times that device \(i\in\mathcal{V}\) transmits to a star center, and as \(n_i^b\in\{0,1\}\) the number of times that device \(i\in\mathcal{V}\) acts as the star center. Note that \(n_i^c\ge 1\) if \(n_i^b=0\). All transmitting devices in a given slot transmit a sparsified version of their error-compensated local parameters \(\mv\theta_j^{(t)}+\mv e_j^{(t-1)}\) over the \(\lfloor\Myfrac{N}{M}\rfloor\) channel uses in that slot. We make here the standard assumption of large training models satisfying \(\lfloor\Myfrac{N}{M}\rfloor<d\) \cite{amiri19fading}.

The transmitted signal of device \(j\in\mathcal{V}\) to a scheduled receiving device \(i\in\mathcal{N}_j\) is expressed as 
\begin{align}
\mv x_{ji}^{(t)} = \Myfrac{\sqrt{\gamma^{(t)}}}{h_{ji}^{\prime(t)}}\mv A\ \mathrm{sparse}_k(\mv\theta_j^{(t)}+\mv e_j^{(t-1)}), \label{eq:received analog signal}
\end{align} 
where \(\mathrm{sparse}_k(\cdot)\) is a sparsifying operator that sets all but the $k$ elements of \(\mv\theta_j^{(t)}+\mv e_j^{(t-1)}\) with the largest absolute values to zero; \(\mv A\in\mb{R}^{\lfloor\Myfrac{N}{M}\rfloor\times d}\) is a compression matrix that adapts the dimension $d$ of parameter vectors to the number of channel uses per slot;  \(h_{ji}^{\prime(t)}\triangleq\sqrt{A_0}(\Myfrac{d_0}{d_{ij}})^{\frac{\gamma}{2}}h_{ji}^{(t)}\) is the effective channel coefficient; and \(\gamma^{(t)}\) is a power scaling factor. 
Assuming an equal power allocation for all slots, the scaling factor \(\gamma^{(t)}\) is chosen to satisfy the average power constraint for any device \(j\in\mathcal{V}\)
by imposing the following inequality
\begin{align}
\gamma^{(t)}\le\Big(\|\mv\phi_j^{(t)}\|^2\sum_{i\in\mathcal{N}_j^{ o}}\vert h_{ij}^{\prime}(t)\vert^{-2}\Big)^{-1}\bar PN\frac{n_j^c}{n_j^{c}+n_j^{ b}}, \label{eq:power scaling factor}
\end{align} where \(\mv\phi_j^{(t)}\triangleq\mv A\ \mathrm{sparse}_k(\mv\theta_j^{(t)}+\mv e_j^{(t-1)})\); and \(\mathcal{N}_j^{ o}\subseteq\mathcal{N}_j\) is defined as a subset of device $j$'s neighbors that are scheduled as the star center. Computing $\gamma^{(t)}$ requires a low-overhead consensus mechanism for the nodes to agree on the minimum for all \(j\in\mathcal{V}\) in the right-hand side of \eqref{eq:power scaling factor} (see e.g., \cite{iutzeler12max-consensus}). 

As a result, the scheduled receiver device \(i\in\mathcal{V}\) in a given slot receives the following signal
\begin{align}
\mv y_i^{(t)} = \sqrt{\gamma^{(t)}}\mv A\sum\limits_{j\in\mathcal{N}_i^{(m)}}\mathrm{sparse}_k(\mv\theta_j^{(t)}+\mv e_j^{(t)}) + \mv n_i^{(t)}. \label{eq:analog received signal}
\end{align}
Upon scaling the received signal \(\mv y_i^{(t)}\) by  \(\Myfrac{1}{\sqrt{\gamma^{(t)}}}\), device \(i\in\mathcal{V}\) estimates the vector \(\sum_{j\in\mathcal{N}_i}\mathrm{sparse}_k(\mv\theta_j^{(t)}+\mv e_j^{(t)})\) by leveraging a compressive sensing algorithm \(f_{\mv A}(\cdot)\), such as LASSO\cite{meinshausen09lasso}. The resulting consensus update for device \(i\in\mathcal{V}\) is given as (cf.~\eqref{eq:consensus udpates})
\begin{multline}
\mv\theta_i^{(t+1)}=w_{ii}\mv\theta_i^{(t)}+\sum\limits_{j\in\mathcal{N}_i}w_{ij}f_{\mv A}\left(\Myfrac{\Re\{\mv y_i^{(t)}\}}{\sqrt{\gamma^{(t)}}}\right)-\\
\eta^{(t)}\hat\nabla f_i(\mv\theta_i^{(t-1)}). \label{eq:consensus udpates for A-DSGD}
\end{multline}
Finally, the accumulated error at device \(i\in\mathcal{V}\) with \(\mv e_i^{(0)} = \mv 0\) is updated similarly to the digital implementation as
\begin{align}
\mv e_i^{(t)}=\mv e_i^{(t-1)}+\left(\mv\theta_i^{(t)}
-\mathrm{sparse}_k(\mv\theta_i^{(t)}+\mv e_i^{(t-1)})\right).\label{eq:accumulated compressed error}
\end{align}


\section{Experiments}\label{sec:Experiments}
In this section, we evaluate the performance of the proposed digital and analog implementations of the DSGD algorithm in a wireless setup with $K=8$ devices. We consider the learning task of classifying $10$ classes of fashion articles over the standard Fashion-MNIST dataset, which is divided into $60,000$ training data samples ($6,000$ samples for each class) and $10,000$ test data samples with each of them corresponding to a \(28\times28\) image. Each device has training data for all classes excluding a uniformly random-selected number (between $2$ and $4$) of classes, with an equal number of samples for each of the available classes. The samples are drawn randomly (without replacement) from the training set. All devices train a common classification model consisting of one input layer and one softmax output layer, yielding a total number $d=28\times 28\times 10 + 10=7850$ of training parameters. 

The connectivity graph (cf.~Fig \ref{subfig:illustration of vertex coloring}) is generated as follows. First, we fix a base star-topology graph with node $1$ located in the center and all other $K-1$ nodes in locations that are specified by a random distance \(d_{i0}\) uniformly distributed in the interval \((20, 200]\)m and a uniformly distributed angle. Next, {\color{black} an edge between each pair of nodes in \(\mathcal{V}\setminus\{1\}\) is independently added to this base graph with probability $p$}. The accuracy of the training model is assessed at each device \(i\in\mathcal{V}\) over the test samples by averaging over $5$ episodes of training. The simulation parameters are set as follows unless otherwise specified. The mini-batch size is set as $\vert\mathcal{D}_i^{(t)}\vert=32$; the communication interval is set as \(\tau = 10\); the path loss parameter is set as \(A_0 =10^{-3.35}\) with \(d_0=1\) m and \(\gamma =3.76\); the noise power is set as \(N_0 = -169\) dBm; \(\bar P =1\) mW is set such that the average received signal-to-noise ratio (SNR) over $K$ devices is around $40$ dB; the total number of channel uses is set as \(N = 30000\); and \(k =d(1-0.4^{\Myfrac{1}{K}})\)  is set for the sparsifying operator \(\mathrm{sparse}_k(\cdot)\) at each device.

As benchmarks, we consider the standard DSGD algorithm that assumes {\color{black} noiseless and interference-free communication} among devices (``ideal communications''), as well as TDMA-based scheduling policies that select only one device for each time slot as the transmitter or the receiver for digital and analog implementations, respectively. We also include a benchmark scheme that executes local SGD updates independently at each device without any communications among devices (``no communications''). 

\begin{figure}[htp]
	\centering
	\includegraphics[width=3.48in]{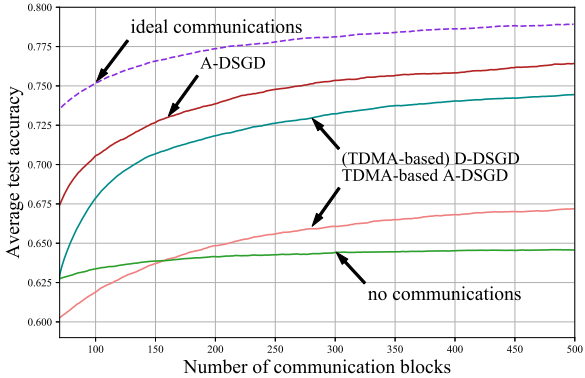}
	\caption{Average test accuracy versus the number of communication blocks with $p=0.1$.}\label{fig:star extended with small prob.}
	\vspace{-.12in}
\end{figure}

\begin{figure}[htp]
	\centering
	\includegraphics[width=3.48in]{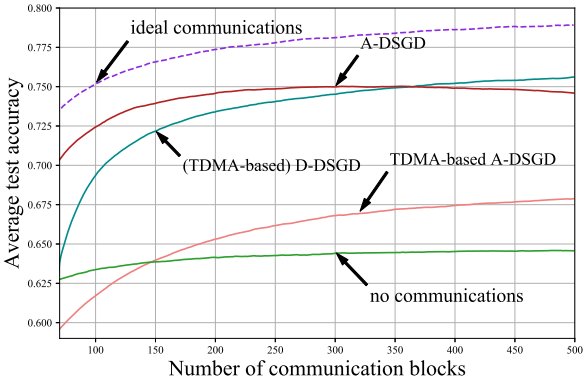}
	\caption{Average test accuracy versus the number of communication blocks with $p=0.3$.}\label{fig:star extended with large prob.}
	\vspace{-.12in}
\end{figure}

The average test accuracy over devices is evaluated versus the number of communication blocks at iteration \(t=\tau, 2\tau, \ldots\) for $p=0.1$ and $p=0.3$ in Figs. \ref{fig:star extended with small prob.} and \ref{fig:star extended with large prob.}, respectively. ``No communications'' is seen to perform poorly due to missing classes and small size of training sets at each local device, and communication is generally beneficial. For small values of $p$, for which the topology closely resembles a star, the analog implementation (``A-DSGD'') is seen to be superior to the digital implementation (``D-DSGD'') from the perspective of both convergence speed and final accuracy. This performance advantage is, nevertheless, not necessarily preserved with larger values of $p$. This is because in a more densely connected topology, the total number $M$ of slots scheduled within one communication block for analog communication gets larger, yielding less observations for the estimation of the sum of parameter vectors (cf.~\eqref{eq:received analog signal}).
TDMA-based schemes are also seen to be generally strongly suboptimal for analog transmission. However, this is not the case for digital transmission, since in this example, the interference-avoidance scheduling scheme adopted by the digital implementation yields the same number of slots per communication block as TDMA.  

\section{Conclusions and Discussions}\label{sec:Conclusion and Discussions}
{\color{black}In this paper, we proposed digital and analog implementations of decentralized FL over wireless D2D networks. As general conclusion, we have observed that over-the-air computing can outperform conventional digital implementations of the DSGD only in star-like topologies. Furthermore, the results highlight the importance of scheduling optimization. Among a few directions for future work, we mention the implementation of gradient-tracking schemes.}

\bibliographystyle{IEEEtran}
\bibliography{DL_ref}

\begin{thebibliography}{10}
\providecommand{\url}[1]{#1}
\csname url@samestyle\endcsname
\providecommand{\newblock}{\relax}
\providecommand{\bibinfo}[2]{#2}
\providecommand{\BIBentrySTDinterwordspacing}{\spaceskip=0pt\relax}
\providecommand{\BIBentryALTinterwordstretchfactor}{4}
\providecommand{\BIBentryALTinterwordspacing}{\spaceskip=\fontdimen2\font plus
\BIBentryALTinterwordstretchfactor\fontdimen3\font minus
  \fontdimen4\font\relax}
\providecommand{\BIBforeignlanguage}[2]{{%
\expandafter\ifx\csname l@#1\endcsname\relax
\typeout{** WARNING: IEEEtran.bst: No hyphenation pattern has been}%
\typeout{** loaded for the language `#1'. Using the pattern for}%
\typeout{** the default language instead.}%
\else
\language=\csname l@#1\endcsname
\fi
#2}}
\providecommand{\BIBdecl}{\relax}
\BIBdecl

\bibitem{savazzi20decentralized}
S.~Savazzi, M.~Nicoli, and V.~Rampa, ``Federated learning with cooperating
  devices: A consensus approach for massive {IoT} networks,'' \emph{to appear
  in IEEE Internet Things J.}, 2020.

\bibitem{chang20distributed}
T.-H. Chang, M.~Hong, H.-T. Wai, X.~Zhang, and S.~Lu, ``Distributed learning in
  the non-convex world: from batch to streaming data, and beyond,'' \emph{arXiv
  preprint arXiv:2001.04786}, 2020.

\bibitem{bekkerman11scaling}
R.~Bekkerman, M.~Bilenko, and J.~Langford, \emph{Scaling up machine learning:
  Parallel and distributed approaches}.\hskip 1em plus 0.5em minus 0.4em\relax
  Cambridge Univ. Press, 2011.

\bibitem{kairouz19advances}
P.~Kairouz, H.~B. McMahan, B.~Avent, A.~Bellet \emph{et~al.}, ``Advances and
  open problems in federated learning,'' \emph{arXiv preprint
  arXiv:1912.04977}, 2019.

\bibitem{ram10distributed}
S.~S. Ram, A.~Nedi{\'c}, and V.~V. Veeravalli, ``Distributed stochastic
  subgradient projection algorithms for convex optimization,'' \emph{Journal of
  optimization theory and applications}, vol. 147, no.~3, pp. 516--545, Dec.
  2010.

\bibitem{xin19decentralized}
R.~Xin, S.~Kar, and U.~A. Khan, ``An introduction to decentralized stochastic
  optimization with gradient tracking,'' \emph{arXiv preprint
  arXiv:1907.09648v2}, 2019.

\bibitem{wang19MATCHA}
J.~Wang, A.~K. Sahu, Z.~Yang, G.~Joshi, and S.~Kar, ``{MATCHA}: Speeding up
  decentralized {SGD} via matching decomposition sampling,'' \emph{arXiv
  preprint arXiv:1905.09435}, 2019.

\bibitem{basu19qsparse}
D.~Basu, D.~Data, C.~Karakus, and S.~Diggavi, ``Qsparse-local-sgd: Distributed
  sgd with quantization, sparsification and local computations,'' in
  \emph{Proc. Advances in Neural Information Processing Systems (NeurIPS)},
  Vancouver, Canada, Dec. 2019.

\bibitem{singh19sparq-sgd}
N.~Singh, D.~Data, J.~George, and S.~Diggavi, ``Sparq-sgd: Event-triggered and
  compressed communication in decentralized stochastic optimization,''
  \emph{arXiv preprint arXiv:1910.14280}, 2019.

\bibitem{chen19joint}
M.~Chen, Z.~Yang, W.~Saad, C.~Yin, H.~V. Poor, and S.~Cui, ``A joint learning
  and communications framework for federated learning over wireless networks,''
  \emph{arXiv preprint arXiv:1909.07972}, 2019.

\bibitem{Gunduz19over-the-air}
M.~M. Amiri and D.~G{\"{u}}nd{\"{u}}z, ``Machine learning at the wireless edge:
  Distributed stochastic gradient descent over-the-air,'' in \emph{Proc. IEEE
  International Symposium on Information Theory (ISIT)}, Paris, France, Jul.
  2019.

\bibitem{zhu20one-bit}
G.~Zhu, Y.~Du, D.~Gunduz, and K.~Huang, ``One-bit over-the-air aggregation for
  communication-efficient federated edge learning: Design and convergence
  analysis,'' \emph{arXiv preprint arXiv:2001.05713}, 2020.

\bibitem{ahn19distillation}
J.-H. Ahn, O.~Simeone, and J.~Kang, ``Wireless federated distillation for
  distributed edge learning with heterogeneous data,'' in \emph{Proc. IEEE
  International Symposium on Personal, Indoor and Mobile Radio Communications
  (PIMRC)}, Istanbul, Turkey, Sep. 2019.

\bibitem{xiao04fast}
L.~Xiao and S.~Boyd, ``Fast linear iterations for distributed averaging,''
  \emph{Systems \& Control Letters}, vol.~53, no.~1, pp. 65--78, Sept. 2004.

\bibitem{amiri19fading}
M.~M. Amiri and D.~Gunduz, ``Federated learning over wireless fading
  channels,'' \emph{to appear in IEEE Trans. Wireless Commun.}, 2019.

\bibitem{Husfeldt15graph}
T.~Husfeldt, ``Graph colouring algorithms,'' 2015.

\bibitem{iutzeler12max-consensus}
F.~Iutzeler, P.~Ciblat, and J.~Jakubowicz, ``Analysis of max-consensus
  algorithms in wireless channels,'' \emph{{IEEE} Trans. Signal Process.},
  vol.~60, no.~11, pp. 6103--6107, Nov. 2012.

\bibitem{meinshausen09lasso}
N.~Meinshausen, B.~Yu \emph{et~al.}, ``Lasso-type recovery of sparse
  representations for high-dimensional data,'' \emph{The annals of statistics},
  vol.~37, no.~1, pp. 246--270, Jan. 2009.

\end{thebibliography}
\end{document}